\documentclass[11pt]{article}

\usepackage[T1]{fontenc}
\usepackage[utf8]{inputenc}
\usepackage{lmodern}
\usepackage{geometry}
\usepackage{amsmath,amssymb}
\usepackage{graphicx}
\usepackage{booktabs}
\usepackage{tabularx}
\usepackage{array}
\usepackage{enumitem}
\usepackage{caption}
\usepackage{float}
\usepackage{placeins}
\usepackage{xurl}
\usepackage[hidelinks]{hyperref}
\usepackage[numbers,sort&compress]{natbib}
\usepackage{microtype}
\usepackage{algorithm}
\usepackage{algpseudocode}

\newcommand{\sys}{\textsc{RL Developer Memory}}

\newcommand{\clip}{\operatorname{clip}}

\newcommand{\argmax}{\operatorname*{arg\,max}}
\newcolumntype{Y}{>{\raggedright\arraybackslash}X}
\newcommand{\cmark}{$\checkmark$}
\newcommand{\xmark}{$\times$}
\algnewcommand{\Input}{\State \textbf{Input:} }
\algnewcommand{\Output}{\State \textbf{Output:} }

\title{Feedback-Normalized Developer Memory for Reinforcement-Learning Coding Agents: A Safety-Gated MCP Architecture}
\author{Mehmet Iscan\textsuperscript{1,*}\\
\small \textsuperscript{1}PythaLab, Yildiz Technical University, Istanbul, T\"urkiye\\
\small \textsuperscript{*}Corresponding author: \href{mailto:miscan@yildiz.edu.tr}{miscan@yildiz.edu.tr}}
\date{May 2026}

\begin{document}
\maketitle

\begin{abstract}
Large language model (LLM) coding agents increasingly interact with repositories, terminals, tests, and execution traces over long software-engineering episodes. Persistent memory is therefore useful, but a static vector store or generic retrieval-augmented generation (RAG) layer is not sufficient for reinforcement-learning (RL) code development, where small implementation details can change Bellman targets, terminal masks, gradient-flow boundaries, target-network updates, or validation claims. This paper presents \sys, a local-first, Model Context Protocol (MCP)-native developer-memory architecture for RL coding agents. The system treats memory selection as a logged contextual decision process: \texttt{issue\_match} emits ranked candidates and telemetry, \texttt{issue\_feedback} normalizes raw labels into bounded canonical rewards, and \texttt{issue\_record\_resolution} links later verified resolutions back to earlier retrieval events. A deterministic control path remains the deployed policy; a diagonal contextual-bandit residual policy runs in shadow mode, logs propensities, and is eligible for active canary only through conservative off-policy-evaluation (OPE) gates. RL/control memories additionally require theory-to-code metadata, validation-tier promotion checks, and review-gated governance.

The implementation is evaluated on a deterministic 200-case RL developer-memory benchmark with RL algorithm bugs, hard negatives, review-gated RL/control cases, and low-risk developer-memory failures. To avoid presenting an implementation revision as a method baseline, the paper separates two evidence layers. First, the same-commit controlled comparison shows that the deterministic control path and the full shadow/OPE configuration both achieve 80.0\% expected-decision accuracy and 100.0\% hard-negative suppression, while the full configuration adds contextual-statistics updates and OPE/shadow telemetry rather than a demonstrated accuracy gain. Second, a patch-replay audit over paired cases reports favorable raw deltas in offline pass rate and recall-delta metrics, but the claim gate does not elevate these deltas to method-superiority claims. Static validation passed 11/11 checks and dynamic integration passed 10/10 cases. The same evidence reports constraints: active learned-policy deployment is unsupported, official-client MCP interoperability is unsupported, p95 latency regresses in live full configuration, and 40 residual failures remain in non-RL path, scope, and alias compatibility cases. The contribution is therefore an auditable memory-control architecture with component-level evidence and explicit claim boundaries, not a universal coding-agent improvement claim.
\end{abstract}


\section{Introduction}

LLM-based coding systems have moved from isolated code completion toward agentic software engineering. Contemporary agents read repository files, issue shell commands, run tests, inspect stack traces, and revise patches over multi-turn trajectories. SWE-bench formalized this shift by evaluating models on real GitHub issues whose solutions require repository understanding, execution evidence, and multi-file coordination \citep{jimenez2024swebench}. SWE-agent further showed that agent-computer interfaces matter because coding agents need structured ways to navigate files, edit code, and execute programs \citep{yang2024sweagent}. ReAct and iterative self-refinement likewise established that useful agent behavior is not a single generation event, but a loop over reasoning, acting, observing, and revising \citep{yao2023react,madaan2023selfrefine}. Once coding agents operate in this regime, memory becomes a software-engineering concern: the agent must preserve prior fixes, rejected hypotheses, terminal outputs, package constraints, path conventions, validation commands, and project-specific decisions across turns and sessions.

The direct response to this requirement is usually to increase the context window or attach retrieval. These mechanisms are important, but they do not by themselves define developer memory. RAG introduced a principled way to condition generation on retrieved non-parametric evidence \citep{lewis2020rag}, and long-term memory systems such as MemoryBank showed that persistent memories can support continuity across interactions \citep{zhong2024memorybank}. Recent self-evolving memory work goes further by treating episodic memory as a utility-bearing object that can be updated from runtime feedback rather than as a passive semantic cache \citep{zhang2026memrl}. Long-term memory benchmarks, however, show that sustained interaction requires more than recall: systems must handle knowledge updates, temporal reasoning, multi-session evidence, and abstention \citep{wu2024longmemeval}. Long-context software-engineering benchmarks similarly emphasize multi-turn tool use and a comprehension-efficiency trade-off in realistic code tasks \citep{qiu2025locobenchagent}. Long-context models also exhibit utilization failures when relevant information is buried among distractors \citep{liu2024lostmiddle}. For code agents, this limitation is more than a recall problem. A retrieved memory can change a command, patch, test oracle, or validation claim; therefore memory selection is a correctness-relevant decision, not merely prompt decoration.

Developer memory is narrower and stricter than general conversational memory. Coding agents work with repository-specific APIs, runtime traces, build systems, file paths, stack frames, and tests. Empirical work on domain-specific code generation shows that general-purpose LLMs underperform when precise domain APIs and structural cues are missing, and that explicit API knowledge can materially affect generated code quality \citep{gu2025domaincode}. Repository-aware graph retrieval, such as graph retrieval augmented code generation, improves the organization of code context, yet it remains retrieval-centric unless the system also evaluates whether retrieved evidence was useful for the downstream repair decision \citep{fedorov2025gracg}. This distinction motivates a developer-memory layer that tracks not only what was similar, but what was safe, verified, rejected, stale, or causally linked to a later resolution.

The need for such a layer becomes sharper in reinforcement-learning code. RL programs are mathematically fragile because small implementation choices can alter the effective objective, credit assignment, or evaluation protocol. Deep RL reproducibility studies show that results are sensitive to seeds, evaluation practice, and reporting choices \citep{henderson2018deep}. Hyperparameter and sensitivity analyses further show that tuning choices, algorithmic variants, and normalization details can substantially change observed performance \citep{eimer2023hyperparameters,adkins2024sensitivity}. Empirical studies of RL program bugs find that failures often arise from algorithmic logic and agent-environment interaction rather than ordinary API misuse \citep{song2026bugs}. A superficially similar memory can therefore be dangerous: a DQN terminal-mask issue, a PPO ratio/clipping error, a SAC entropy-temperature sign error, or a target-network detach bug may share words with a prior fix while requiring a different theory-to-code obligation.

Existing memory and self-correction systems provide useful ingredients but do not close this control gap. Reflexion stores verbal feedback for later decisions \citep{shinn2023reflexion}; interactive debugging systems combine reinforcement learning, developer feedback, and memory for code repair \citep{krishnamoorthy2025debugging}; and self-correction methods can improve multi-turn repair behavior \citep{madaan2023selfrefine,kumar2024selfcorrect}. These systems primarily optimize generation or repair trajectories. They do not require every memory access to be logged as a contextual decision, normalize heterogeneous developer feedback into a bounded reward contract, link delayed verified resolutions back to specific retrieval events, or gate learned memory reranking through conservative OPE before deployment. At the same time, active online exploration is risky in developer environments, which motivates shadow execution and offline policy evaluation rather than immediate replacement of a deterministic retrieval baseline \citep{gassert2024shadow,guissouma2023shadow,dudik2011doubly,swaminathan2015counterfactual}.

Persistent developer memory also introduces safety and governance risk. Code models can memorize and expose sensitive training data \citep{alkaswan2024traces}; deployed code helpers may mishandle credentials or project metadata \citep{sheggam2024security}; long-term assistant memory raises retention, access-control, and consent concerns \citep{lee2024ethicalmemory}; and agentic tool use expands the attack surface for prompt injection and data manipulation \citep{khan2024agenticsecurity}. The Model Context Protocol (MCP) provides a standardized interface through which servers expose tools, resources, and prompts to clients \citep{mcpdocs2026}. MCP is therefore a natural boundary for developer-memory tools, but the protocol boundary alone is not a safety mechanism. A memory server that returns semantically similar records without feedback normalization, delayed-credit links, and review gates can still reinforce stale or false-positive fixes.

This paper introduces \sys, a feedback-normalized, delayed-reward-aware, contextual-bandit-shadowed, OPE-gated, theory-to-code-traceable developer-memory architecture for RL coding agents. The system is implemented as a local-first MCP server. A deterministic ranker remains the deployed baseline; a contextual-bandit residual policy runs in shadow mode and logs propensities; raw developer feedback is converted into a bounded canonical reward vocabulary; later verified resolutions are linked to earlier retrieval events; conservative OPE gates any active canary behavior; and RL/control memories carry metadata that binds algorithmic obligations to code anchors, validation tiers, and review state. The goal is not to claim a new LLM, a new RL algorithm, or universal coding-agent improvement. The goal is to make developer memory auditable, trainable under delayed feedback, and safe enough to evaluate component by component.

\paragraph{Contributions.}
The paper makes the following contributions.
\begin{enumerate}[leftmargin=2em]
    \item It formalizes developer memory for RL coding agents as a logged contextual decision process with feedback normalization, delayed reward linking, and conservative residual policy learning.
    \item It describes a local-first MCP architecture whose matching, feedback, and resolution-recording tools persist retrieval events, ranked candidates, canonical feedback, delayed reward links, bandit decisions, and OPE summaries.
    \item It introduces RL/control memory governance through theory-to-code metadata, validation-tier promotion, audit findings, review-gated consolidation, and traceable code anchors.
    \item It reports a component-level evaluation on the provided deterministic 200-case benchmark. The evaluation reports measured favorable deltas, unchanged telemetry rates, latency regressions, and residual failure families without claiming broad end-to-end coding-agent improvement.
\end{enumerate}

\section{Method}

\subsection{Framework overview}

\begin{figure}[t]
\centering
\includegraphics[width=\linewidth]{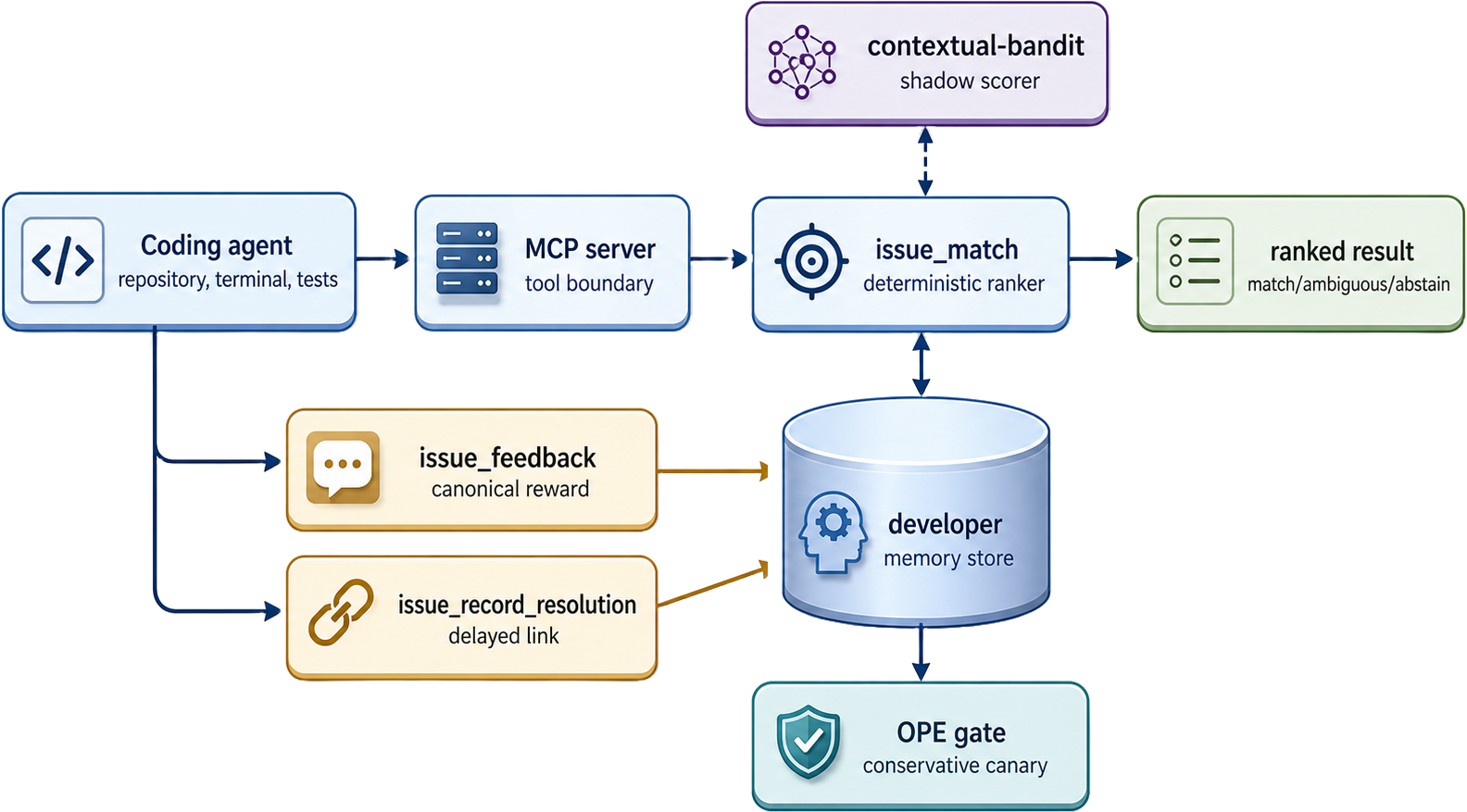}
\caption{System architecture. The deterministic ranker remains the deployed baseline; the contextual bandit runs in shadow mode and only supplies logged propensities and bounded residual scores unless the OPE gate authorizes a conservative low-risk canary.}
\label{fig:architecture}
\end{figure}

\sys\ is a local-first MCP server that interposes a memory-control layer between a coding agent and a persistent developer-memory store. Figure~\ref{fig:architecture} summarizes the architecture. The live retrieval path begins with \texttt{issue\_match}. The query is normalized into a structured profile, candidate memories are retrieved, a deterministic ranker orders them, and a decision policy returns \texttt{match}, \texttt{ambiguous}, or \texttt{abstain}. In parallel, a contextual-bandit residual policy scores the same candidates in shadow mode. The shadow policy does not alter the deployed rank unless a separate OPE gate later authorizes active canary behavior. The update path is split between \texttt{issue\_feedback}, which records explicit user feedback, and \texttt{issue\_record\_resolution}, which stores a verified fix and links the delayed outcome to an earlier retrieval event.

The architecture is feedback-normalized because raw labels are mapped into a bounded canonical reward contract before persistence. It is delayed-reward-aware because verified resolutions are linked back to retrieval events. It is contextual-bandit-shadowed because learned utility estimates are collected beside the deterministic baseline without unverified intervention. It is OPE-gated because active rollout is blocked unless logged evidence supports conservative policy selection. It is theory-to-code traceable because RL/control memories must carry metadata connecting algorithmic obligations to code anchors and validation tiers.

\subsection{Problem definition}

Consider a coding episode indexed by time $t$. The agent observes a developer context
\begin{equation}
    x_t = (e_t, q_t, p_t, b_t, c_t, u_t),
    \label{eq:context}
\end{equation}
where $e_t$ is an error or symptom, $q_t$ is the current command or natural-language query, $p_t$ is project scope, $b_t$ is repository and path evidence, $c_t$ is execution context such as stack traces or environment metadata, and $u_t$ is the user/session scope. The memory bank is
\begin{equation}
    \mathcal{M}_t = \{m_i\}_{i=1}^{N_t}, \qquad
    m_i = (h_i, v_i, a_i, g_i),
    \label{eq:memory-bank}
\end{equation}
where $h_i$ is the canonical failure pattern, $v_i$ is a variant/fix record, $a_i$ is audit and validation metadata, and $g_i$ is feedback/governance state.

The memory system must select an action from
\begin{equation}
    \mathcal{A}(x_t,\mathcal{M}_t)=\{\texttt{match}(m_i),\texttt{ambiguous},\texttt{abstain}\}.
    \label{eq:action-space}
\end{equation}
A match action returns one or more memories to the coding agent. The realized utility is not always observed immediately. It may appear after a later test run, verified patch, or resolution write-back. Therefore, for a memory decision $a_t$, the relevant outcome is a delayed signal
\begin{equation}
    r_{t+\Delta} \in [-1,1], \qquad \Delta \geq 0,
    \label{eq:delayed-reward}
\end{equation}
associated with a retrieval event rather than necessarily with the same tool call.

The objective is not to maximize unconstrained online reward. In RL code development, unsafe exploration can produce incorrect patches, stale code paths, or invalid algorithmic claims. The objective is a constrained memory-control problem:
\begin{equation}
\begin{aligned}
    \max_{\pi}\quad & V(\pi)=\mathbb{E}_{x\sim \mathcal{D}}\,\mathbb{E}_{a\sim\pi(\cdot\mid x)}[r(x,a)] \\
    \text{s.t.}\quad & \Pr_{\pi}(\mathrm{false\ positive}) \leq \rho_{\max}, \\
    & L_{95}(\pi) \leq L_{\max}, \\
    & \pi \text{ is supported by logged evidence and OPE}, \\
    & \pi \text{ cannot override RL/control review gates}.
\end{aligned}
\label{eq:constrained-objective}
\end{equation}
This formulation follows the practical constraints implied by reproducibility and sensitivity results in RL \citep{henderson2018deep,eimer2023hyperparameters,adkins2024sensitivity}, the safety requirement for shadow learning \citep{gassert2024shadow,guissouma2023shadow}, and the need to evaluate policies from logged feedback \citep{dudik2011doubly,swaminathan2015counterfactual}.

\subsection{Input normalization and candidate features}

The raw context is normalized into a query profile
\begin{equation}
    z_t = \psi(x_t) = (F_t, R_t, E_t, K_t, P_t, S_t),
    \label{eq:query-profile}
\end{equation}
where $F_t$ is the inferred error family, $R_t$ is the root-cause class, $E_t$ is a set of entity slots, $K_t$ is a token/signature vector, $P_t$ is project and repository scope, and $S_t$ contains RL/control hints such as algorithm family, runtime stage, validation tier, and theory-claim type.

For each candidate memory $m_i$, the system computes a bounded feature vector
\begin{equation}
    \boldsymbol{\phi}_{t,i}=\phi(z_t,m_i)\in[-1,1]^d.
    \label{eq:feature-vector}
\end{equation}
The implementation uses lexical, dense, exception, command, path, entity, scope, family, root-cause, feedback, success-prior, negative-applicability, session, memory-kind, problem-family, algorithm-family, theory, and validation-tier features. Two examples clarify the symbolic form. Recency is represented as
\begin{equation}
    \phi_{\mathrm{rec}}(m_i)=\frac{1}{1+d_i/30},
    \label{eq:recency-feature}
\end{equation}
where $d_i$ is the age in days. Feedback-derived quality is represented as
\begin{equation}
    \phi_{\mathrm{fb}}(m_i)=\clip_{[0,1]}\left(0.42\,c_i+0.33\,s_i+0.25\,u_i-0.22\,\rho_i\right),
    \label{eq:feedback-feature}
\end{equation}
where $c_i$ is confidence, $s_i$ is memory strength, $u_i$ is smoothed success ratio, and $\rho_i$ is rejection ratio.

\subsection{Deterministic retrieval and decision surface}

The first-stage retriever produces a candidate set
\begin{equation}
    \mathcal{C}_t=\mathcal{R}(z_t,\mathcal{M}_t,k), \qquad \mathcal{C}_t\subseteq \mathcal{M}_t.
    \label{eq:candidate-set}
\end{equation}
A deterministic weighted ranker then computes
\begin{equation}
    s_0(z_t,m_i)=\clip_{[0,0.999]}\left(\boldsymbol{w}^{\top}\boldsymbol{\phi}_{t,i}\right).
    \label{eq:deterministic-score}
\end{equation}
Let $m_{(1)}$ and $m_{(2)}$ be the top two candidates after deterministic sorting. The margin is
\begin{equation}
    \delta_t=s_0(z_t,m_{(1)})-s_0(z_t,m_{(2)}),
    \label{eq:margin}
\end{equation}
with $s_0(z_t,m_{(2)})=0$ when no second candidate exists. The decision rule is
\begin{equation}
D_t=
\begin{cases}
\texttt{match}, & s_0(z_t,m_{(1)})\geq \tau_a \land \delta_t\geq \tau_m \land \sigma(z_t,m_{(1)})=1,\\
\texttt{ambiguous}, & s_0(z_t,m_{(1)})\geq \tau_w \land \delta_t<\tau_m,\\
\texttt{abstain}, & \mathcal{C}_t=\varnothing \lor s_0(z_t,m_{(1)})<\tau_w \lor \sigma(z_t,m_{(1)})=0.
\end{cases}
\label{eq:decision-rule}
\end{equation}
Here $\tau_a$ is the accept threshold, $\tau_w$ is the weak threshold, $\tau_m$ is the ambiguity margin, and $\sigma$ is a specificity predicate for path, command, entity, scope, and domain compatibility. This rule is intentionally conservative: a candidate cannot be promoted solely because of high lexical similarity if specificity evidence is missing.

\subsection{Feedback normalization}

Developer feedback is sparse, delayed, and linguistically inconsistent. The system therefore applies a canonical normalization map
\begin{equation}
    g(y^{\mathrm{raw}}_t,\bar r_t)=\left(y_t,r_t,\ell_t,\boldsymbol{v}_t\right),
    \label{eq:feedback-map}
\end{equation}
where $y_t$ is a canonical feedback type, $r_t\in[-1,1]$ is a bounded reward, $\ell_t\in\{0,1\}$ indicates whether the event is allowed to update learning state, and $\boldsymbol{v}_t$ stores auditable reward metadata. The canonical vocabulary is
\begin{equation}
\begin{aligned}
\mathcal{Y}=\{&\texttt{candidate\_accepted},\texttt{candidate\_rejected},\texttt{fix\_verified},\texttt{false\_positive},\\
&\texttt{merge\_confirmed},\texttt{merge\_rejected},\texttt{split\_confirmed},\texttt{split\_rejected}\}.
\end{aligned}
\label{eq:canonical-vocab}
\end{equation}
The default scalar reward map is
\begin{equation}
\begin{aligned}
    r(\texttt{fix\_verified}) &= 1.00, &
    r(\texttt{false\_positive}) &= -1.00,\\
    r(\texttt{candidate\_accepted}) &= 0.35, &
    r(\texttt{candidate\_rejected}) &= -0.60,\\
    r(\texttt{merge\_confirmed}) &= 0.40, &
    r(\texttt{merge\_rejected}) &= -0.40,\\
    r(\texttt{split\_confirmed}) &= 0.40, &
    r(\texttt{split\_rejected}) &= -0.40.
\end{aligned}
\label{eq:reward-map}
\end{equation}
Aliases such as \texttt{accepted\_helpful}, \texttt{accepted\_unhelpful}, and \texttt{rejected} resolve to the canonical set. The alias \texttt{neutral} maps to zero reward with $\ell_t=0$, preventing non-evaluative labels from becoming negative evidence. Reward normalization is required for stable RL updates because normalization affects effective learning rates and value scales \citep{lyle2024normalization}.

\subsection{Delayed reward linking}

Let $E_t$ denote the retrieval event written by \texttt{issue\_match}. A resolution may later arrive through \texttt{issue\_record\_resolution}. The linker selects the retrieval event
\begin{equation}
E^\star=
\begin{cases}
E^{\mathrm{explicit}}, & \text{if an explicit retrieval-event id is provided and exists},\\
\argmax_{E\in\mathcal{H}_t} \eta(E,z_t), & \text{if a compatible recent session event exists},\\
\varnothing, & \text{otherwise},
\end{cases}
\label{eq:link-event}
\end{equation}
where $\mathcal{H}_t$ is the recent event window and $\eta$ is a compatibility score over session, repository, project scope, and query profile. Link confidence is
\begin{equation}
    \kappa(E^\star)=
    \begin{cases}
        1.00, & E^\star=E^{\mathrm{explicit}},\\
        0.75, & E^\star\in\mathcal{H}_t,\\
        0, & E^\star=\varnothing.
    \end{cases}
    \label{eq:link-confidence}
\end{equation}
Given the top retrieved candidate $m_{(1)}$ and the recorded resolution $z_{t+\Delta}$, the implicit feedback type is
\begin{equation}
\hat y_{t+\Delta}= 
\begin{cases}
\texttt{false\_positive}, & \text{if resolution notes mark the retrieved memory as wrong},\\
\texttt{fix\_verified}, & \text{if } z_{t+\Delta} \text{ matches } m_{(1)}\text{'s pattern/variant},\\
\texttt{candidate\_rejected}, & \text{otherwise}.
\end{cases}
\label{eq:implicit-feedback}
\end{equation}
The delayed feedback event is then normalized through Eq.~\eqref{eq:feedback-map}. The storage layer enforces idempotence by preventing duplicate links for the tuple
\begin{equation}
    \xi=(E^\star,\mathrm{pattern\_id},\mathrm{variant\_id},\hat y_{t+\Delta}).
    \label{eq:idempotence-key}
\end{equation}
This mechanism improves auditability and sample use under delayed feedback, but it remains an approximation of temporal credit assignment; delayed reward is known to create misattribution risk in RL \citep{qu2025latentreward}.

\subsection{Contextual-bandit shadow learning}

Selecting a memory candidate for a query is closer to a single-step contextual recommendation problem than to a fully observed Markov decision process. Contextual bandits are widely used for utility-driven ranking from logged interactions \citep{li2010contextual,gampa2021banditrank}, and memory-augmented contextual bandits are suitable when user feedback is sparse \citep{shen2018memorybandit}. \sys\ uses a diagonal linear upper-confidence residual policy. For each feature dimension $j$, the state variables are $(A_j,b_j,n_j)$ with ridge initialization
\begin{equation}
    A_j^{(0)}=\lambda, \qquad b_j^{(0)}=0, \qquad n_j^{(0)}=0.
    \label{eq:bandit-init}
\end{equation}
The posterior mean estimate is
\begin{equation}
    \hat\theta_j=\frac{b_j}{A_j}.
    \label{eq:theta}
\end{equation}
For candidate $m_i$, the shadow utility estimate and uncertainty are
\begin{equation}
\begin{aligned}
    \mu_{t,i} &= \sum_{j=1}^{d}\hat\theta_j\phi_{t,i,j},\\
    u_{t,i} &= \left(\sum_{j=1}^{d}\frac{\phi_{t,i,j}^{2}}{A_j}\right)^{1/2}.
\end{aligned}
\label{eq:bandit-mean-uncertainty}
\end{equation}
The residual perturbation is capped:
\begin{equation}
    \Delta_{t,i}=\clip_{[-\Delta_{\max},\Delta_{\max}]}
    \left(\gamma(\mu_{t,i}+\alpha u_{t,i})\right),
    \label{eq:residual-delta}
\end{equation}
and the shadow score is
\begin{equation}
    s_b(z_t,m_i)=\clip_{[0,0.999]}\left(s_0(z_t,m_i)+\Delta_{t,i}\right).
    \label{eq:shadow-score}
\end{equation}
The target propensity used for logging is a softmax over shadow scores:
\begin{equation}
    \pi_b(m_i\mid z_t)=\frac{\exp(s_b(z_t,m_i)/T)}{\sum_{m\in\mathcal{C}_t}\exp(s_b(z_t,m)/T)}.
    \label{eq:target-propensity}
\end{equation}
The behavior propensity is deterministic for the live baseline:
\begin{equation}
    p_t(m_i\mid z_t)=\mathbb{I}[i=1].
    \label{eq:behavior-propensity}
\end{equation}
Only feedback-bearing events update the diagonal statistics:
\begin{equation}
\begin{aligned}
    A_j &\leftarrow A_j+\ell_t\kappa_t\phi_{t,a_t,j}^{2},\\
    b_j &\leftarrow b_j+\ell_t\kappa_t r_t\phi_{t,a_t,j},\\
    n_j &\leftarrow n_j+\ell_t\mathbb{I}[|\phi_{t,a_t,j}|>0].
\end{aligned}
\label{eq:bandit-update}
\end{equation}
Thus the learned policy is trained from normalized, confidence-weighted, logged feedback while preserving the deterministic ranker as the deployed controller.

\subsection{OPE-gated conservative rollout}

Because active online exploration is risky in developer environments, \sys\ evaluates candidate learned policies from logged feedback before any active canary. OPE estimates policy value from rows
\begin{equation}
    \mathcal{D}_{\mathrm{log}}=\{(z_i,a_i,p_i,r_i,\pi_b(a_i\mid z_i))\}_{i=1}^{n}.
    \label{eq:logged-data}
\end{equation}
The inverse-propensity estimator is
\begin{equation}
    \widehat{V}_{\mathrm{IPS}}(\pi_b)=\frac{1}{n}\sum_{i=1}^{n}\frac{\pi_b(a_i\mid z_i)}{p_i}r_i.
    \label{eq:ips}
\end{equation}
The self-normalized estimator is
\begin{equation}
    \widehat{V}_{\mathrm{SNIPS}}(\pi_b)=
    \frac{\sum_{i=1}^{n}\frac{\pi_b(a_i\mid z_i)}{p_i}r_i}
    {\sum_{i=1}^{n}\frac{\pi_b(a_i\mid z_i)}{p_i}}.
    \label{eq:snips}
\end{equation}
The doubly robust estimator is
\begin{equation}
    \widehat{V}_{\mathrm{DR}}(\pi_b)=\frac{1}{n}\sum_{i=1}^{n}\left[\hat q(z_i,\pi_b)+\frac{\pi_b(a_i\mid z_i)}{p_i}\left(r_i-\hat q(z_i,a_i)\right)\right],
    \label{eq:dr}
\end{equation}
where $\hat q$ is a bounded reward model derived from calibrated candidate scores. The promotion lower bound is
\begin{equation}
    L_{0.95}(\pi_b)=\operatorname{LCB}_{0.95}\left(\{v_i^{\mathrm{DR}}\}_{i=1}^{n}\right),
    \label{eq:lcb}
\end{equation}
computed by bootstrap over bounded DR contributions. The active gate is
\begin{equation}
\begin{aligned}
G(\pi_b)=\mathbb{I}[&n_{\mathrm{sup}}\geq n_{\min}\ \land\ \rho_{\mathrm{fp}}\leq \rho_{\max}\ \land\ L_{0.95}(\pi_b)>V_0+\epsilon \\
&\land\ L_{95}(\pi_b)\leq L_{\max}\ \land\ \chi_{\mathrm{lowrisk}}=1\ \land\ \chi_{\mathrm{rlcontrol}}=0].
\end{aligned}
\label{eq:active-gate}
\end{equation}
Here $n_{\mathrm{sup}}$ is target-policy support in logged data, $\rho_{\mathrm{fp}}$ is the false-positive rate, $V_0$ is the deterministic baseline proxy, and $\chi_{\mathrm{rlcontrol}}$ blocks active learned override for RL/control memories. Conservative offline RL motivates this kind of lower-bound and support-aware deployment criterion \citep{kumar2020cql,yu2021combo}.

\subsection{Theory-to-code traceability and governance}

RL/control memories require additional structure because a plausible patch may still violate the intended algorithm. Each RL/control memory includes metadata
\begin{equation}
    a_i=(k_i,f_i,\omega_i,\nu_i,\tau_i,\mathcal{A}_i,\mathcal{V}_i,\mathcal{B}_i),
    \label{eq:metadata}
\end{equation}
where $k_i$ is memory kind, $f_i$ is problem family, $\omega_i$ is theory-claim type, $\nu_i$ is validation tier, $\tau_i$ is runtime stage, $\mathcal{A}_i$ is artifact references, $\mathcal{V}_i$ is validation payload, and $\mathcal{B}_i$ is audit findings. A theory-to-code registry maps obligations to anchors:
\begin{equation}
    \Gamma: (o,\Omega,\mathcal{H}) \mapsto \{(\mathrm{file},\mathrm{symbol},\mathrm{line},\mathrm{check})\},
    \label{eq:anchor-map}
\end{equation}
where $o$ is an objective, $\Omega$ is an update equation or theory-level obligation, and $\mathcal{H}$ is an assumption set. For example, a temporal-difference target obligation can be represented as
\begin{equation}
    y_t=r_t+\gamma(1-d_t)\max_{a'}Q_{\bar\theta}(s_{t+1},a'),
    \label{eq:td-target}
\end{equation}
which requires explicit anchors for terminal masking, target-network use, and gradient detachment.

Promotion is not controlled by a requested validation tier alone. The applied tier is a function of validation evidence, audit findings, seeds, artifacts, and review state:
\begin{equation}
    \nu_i^{\mathrm{applied}}=\Pi(\nu_i^{\mathrm{requested}},\mathcal{V}_i,\mathcal{B}_i,\mathcal{A}_i,\mathrm{review}_i).
    \label{eq:promotion}
\end{equation}
Review-gated governance uses the lifecycle
\begin{equation}
    \mathcal{L}=\{\texttt{retain},\texttt{merge},\texttt{split},\texttt{demote},\texttt{review}\},
    \label{eq:lifecycle}
\end{equation}
so that memory entries can be revised or demoted rather than blindly retained.

\subsection{Training and evaluation phases}

The implementation separates training and deployment into four phases. Offline bootstrap replays deterministic benchmark cases and initializes the logged dataset. Online-shadow operation preserves the deterministic baseline while collecting target propensities and normalized feedback. OPE summarizes support, false positives, and conservative lower bounds. Active canary is disabled unless Eq.~\eqref{eq:active-gate} evaluates to one.

The offline and shadow training data after $n$ feedback-bearing events is
\begin{equation}
    \mathcal{D}_n=\{(z_i,\boldsymbol{\phi}_{i,a_i},a_i,p_i,\pi_b(a_i\mid z_i),r_i,\ell_i,\kappa_i)\}_{i=1}^{n}.
    \label{eq:training-data}
\end{equation}
The empirical training loss for the diagonal residual model can be written as ridge-regularized squared reward prediction,
\begin{equation}
    \mathcal{J}(\boldsymbol{\theta})=\sum_{i=1}^{n}\ell_i\kappa_i\left(r_i-\boldsymbol{\theta}^{\top}\boldsymbol{\phi}_{i,a_i}\right)^2+\lambda\lVert\boldsymbol{\theta}\rVert_2^2,
    \label{eq:ridge-loss}
\end{equation}
whose diagonal online solution is implemented by Eq.~\eqref{eq:bandit-update}. The output of training is not a new language model; it is a bounded residual scorer plus an auditable OPE report.

\begin{algorithm}[H]
\caption{Logged memory matching with contextual-bandit shadow scoring}
\label{alg:matching}
\begin{algorithmic}[1]
\Input context $x_t$, memory bank $\mathcal{M}_t$, deterministic weights $\boldsymbol{w}$, bandit state $(\boldsymbol{A},\boldsymbol{b})$, thresholds $(\tau_a,\tau_w,\tau_m)$
\Output decision $D_t$, visible matches $O_t$, retrieval event $E_t$, shadow telemetry $B_t$
\State $z_t \gets \psi(x_t)$ \Comment{normalize error, command, path, scope, stack, entities, and RL hints}
\State $\mathcal{C}_t \gets \mathcal{R}(z_t,\mathcal{M}_t,k)$ \Comment{candidate retrieval}
\ForAll{$m_i\in\mathcal{C}_t$}
    \State $\boldsymbol{\phi}_{t,i}\gets\phi(z_t,m_i)$ using Eq.~\eqref{eq:feature-vector}
    \State $s_0(z_t,m_i)\gets \clip_{[0,0.999]}(\boldsymbol{w}^{\top}\boldsymbol{\phi}_{t,i})$ using Eq.~\eqref{eq:deterministic-score}
\EndFor
\State sort $\mathcal{C}_t$ by $s_0$ and deterministic tie-breakers
\State $D_t\gets$ decision from Eq.~\eqref{eq:decision-rule}
\ForAll{top-$k_b$ candidates $m_i$}
    \State compute $(\mu_{t,i},u_{t,i})$ using Eq.~\eqref{eq:bandit-mean-uncertainty}
    \State compute $\Delta_{t,i}$ and $s_b(z_t,m_i)$ using Eq.~\eqref{eq:residual-delta} and Eq.~\eqref{eq:shadow-score}
    \State compute target propensity $\pi_b(m_i\mid z_t)$ using Eq.~\eqref{eq:target-propensity}
\EndFor
\State $E_t\gets$ persist retrieval event, candidates, deterministic decision, behavior propensity, and shadow telemetry
\State $O_t\gets$ top visible candidates if $D_t\neq\texttt{abstain}$, otherwise $\varnothing$
\State \Return $(D_t,O_t,E_t,B_t)$
\end{algorithmic}
\end{algorithm}

\begin{algorithm}[H]
\caption{Feedback normalization and delayed-reward update}
\label{alg:feedback}
\begin{algorithmic}[1]
\Input raw feedback $y_t^{\mathrm{raw}}$ or resolution $z_{t+\Delta}$, optional retrieval event id, logged events $\mathcal{H}_t$, bandit state $(\boldsymbol{A},\boldsymbol{b})$
\Output feedback event $F_t$, delayed link $L_t$, updated bandit state
\If{explicit feedback is supplied}
    \State $(y_t,r_t,\ell_t,\boldsymbol{v}_t)\gets g(y_t^{\mathrm{raw}},\bar r_t)$ using Eq.~\eqref{eq:feedback-map}
    \State $E^\star\gets$ supplied retrieval event
    \State $\kappa_t\gets1$
\Else
    \State $E^\star\gets$ selected event from Eq.~\eqref{eq:link-event}
    \State $\kappa_t\gets\kappa(E^\star)$ using Eq.~\eqref{eq:link-confidence}
    \State $\hat y_{t+\Delta}\gets$ implicit feedback type from Eq.~\eqref{eq:implicit-feedback}
    \State $(y_t,r_t,\ell_t,\boldsymbol{v}_t)\gets g(\hat y_{t+\Delta},\varnothing)$
\EndIf
\If{$E^\star=\varnothing$ or idempotence key $\xi$ from Eq.~\eqref{eq:idempotence-key} already exists}
    \State \Return no-learning delayed-link record
\EndIf
\State persist feedback event $F_t$ and delayed link $L_t$
\If{$\ell_t=1$}
    \State update $(\boldsymbol{A},\boldsymbol{b})$ using Eq.~\eqref{eq:bandit-update}
\EndIf
\State \Return $(F_t,L_t,\boldsymbol{A},\boldsymbol{b})$
\end{algorithmic}
\end{algorithm}

\begin{algorithm}[H]
\caption{OPE-gated conservative rollout}
\label{alg:ope}
\begin{algorithmic}[1]
\Input logged dataset $\mathcal{D}_{\mathrm{log}}$, target policy $\pi_b$, deterministic baseline proxy $V_0$, latency summary $L_{95}$, safety thresholds $(n_{\min},\rho_{\max},\epsilon,L_{\max})$
\Output rollout recommendation $R\in\{\texttt{hold\_shadow},\texttt{blocked},\texttt{eligible\_for\_canary}\}$
\State compute target support $n_{\mathrm{sup}}$ from rows where $p_i>0$ and target markers are observed
\State compute false-positive rate $\rho_{\mathrm{fp}}$
\State compute $\widehat{V}_{\mathrm{IPS}}$, $\widehat{V}_{\mathrm{SNIPS}}$, and $\widehat{V}_{\mathrm{DR}}$ using Eq.~\eqref{eq:ips}--Eq.~\eqref{eq:dr}
\State compute lower bound $L_{0.95}(\pi_b)$ using Eq.~\eqref{eq:lcb}
\If{$n_{\mathrm{sup}}<n_{\min}$}
    \State \Return \texttt{blocked: insufficient support}
\ElsIf{$\rho_{\mathrm{fp}}>\rho_{\max}$}
    \State \Return \texttt{blocked: false-positive risk}
\ElsIf{$L_{0.95}(\pi_b)\leq V_0+\epsilon$}
    \State \Return \texttt{hold\_shadow}
\ElsIf{$L_{95}>L_{\max}$ or low-risk and RL/control predicates fail}
    \State \Return \texttt{blocked: operational safety}
\Else
    \State \Return \texttt{eligible\_for\_canary}
\EndIf
\end{algorithmic}
\end{algorithm}

\FloatBarrier

\section{Results}

\subsection{Experimental framework and test suite}

The evaluation uses the comparative evidence package supplied with the implementation. The tested framework is a Python local-first MCP developer-memory server backed by a SQLite memory store. The primary software paths are the retrieval layer, deterministic ranker, feedback-normalization contract, delayed-reward linker, contextual-bandit shadow scorer, OPE reporter, RL/control metadata validators, and MCP tool surface. The live MCP evidence used a local stdio JSON-RPC fallback because the official MCP client module was unavailable in the evaluation environment; this supports live subprocess/tool-call execution evidence but not official-client interoperability.

The main correction in this revision is the separation of method baselines from patch-replay evidence. A patch-level delta is an engineering regression audit, not a controlled method baseline. The controlled same-commit comparison therefore uses the measured deterministic control path and the measured full configuration. The requested pure RAG and intermediate ablation variants are shown explicitly as evidence gaps rather than inferred from patch data. This convention prevents a reviewer from reading implementation deltas as ablation evidence.

Table~\ref{tab:test-suite} lists the test levels. Static validation checked compile/import/schema readiness. Dynamic integration executed 10 cases over the live code path. The 200-case benchmark was run in offline deterministic-control, offline full-configuration, online-shadow, live deterministic-control, and live full-configuration modes. The benchmark included 80 hard-negative cases and 200 algorithm-family-labeled cases, with 160 RL-family cases and 40 non-RL developer-memory cases. Patch-replay comparisons preserved case identity and content hashes and are reported only as regression-audit evidence.

\begin{table}[tbp]
\centering
\caption{Evaluation levels and supported claim boundaries.}
\label{tab:test-suite}
\small
\begin{tabularx}{\linewidth}{lrrY}
\toprule
Evaluation level & Cases & Status & Supported evidence \\
\midrule
Static validation & 11 checks & 11/11 pass & Compile/import/schema readiness; not behavioral performance. \\
Dynamic integration & 10 & 10/10 pass & End-to-end tool-chain execution on a small integration set. \\
Offline control/full & 200 + 200 & complete with failures & Same-commit controlled comparison of deterministic control and full shadow/OPE instrumentation. \\
Online-shadow & 200 & complete with failures & Shadow contextual-bandit telemetry and delayed-feedback behavior. \\
Live MCP control/full & 200 + 200 & complete with failures & Live subprocess/tool-call feasibility under local stdio JSON-RPC fallback. \\
Hard-negative subset & 80 & complete & Suppression of known confusing neighbors. \\
Patch-replay audit & 200 & case ids match & Engineering regression deltas only; not method-superiority evidence. \\
\bottomrule
\end{tabularx}
\end{table}

\subsection{Benchmark composition}

\begin{figure}[tbp]
\centering
\includegraphics[width=0.80\linewidth]{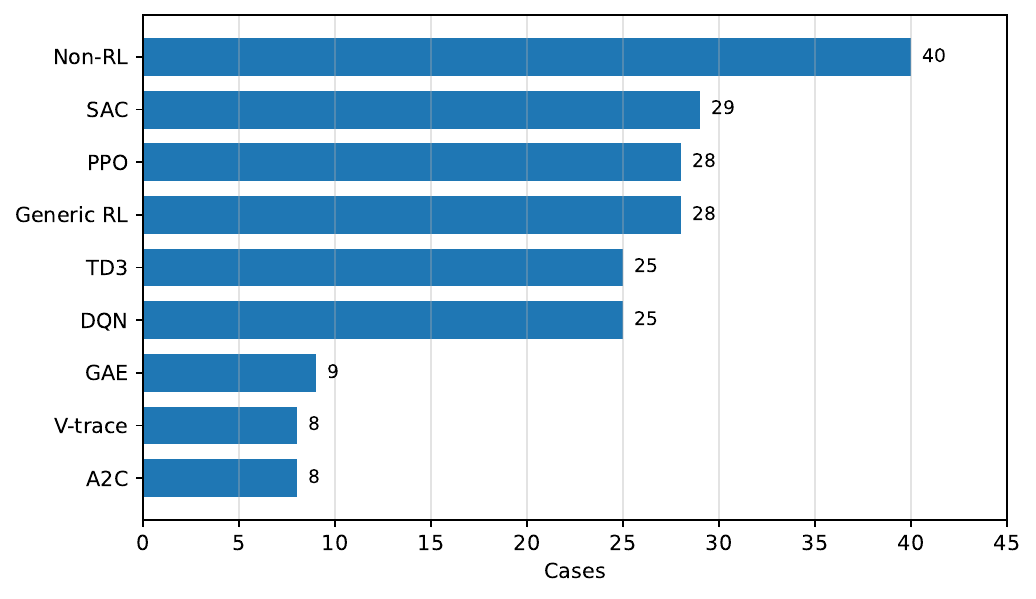}
\caption{Benchmark composition by algorithm family. The 40 non-RL cases are retained because the system is a developer-memory architecture, not only an RL-bug classifier.}
\label{fig:family-composition}
\end{figure}

The benchmark covers several RL algorithm families and a non-RL residual group. Table~\ref{tab:family-composition} and Figure~\ref{fig:family-composition} show the distribution. The RL-related cases cover DQN, PPO, SAC, TD3, GAE, A2C, V-trace, and generic RL families. The non-RL group contains low-risk developer-memory failures such as project-scope and path/alias issues.

\begin{table}[tbp]
\centering
\caption{Benchmark composition by algorithm family.}
\label{tab:family-composition}
\small
\begin{tabular}{lr@{\hspace{2em}}lr}
\toprule
Family & Cases & Family & Cases \\
\midrule
A2C & 8 & DQN & 25 \\
GAE & 9 & Generic RL & 28 \\
PPO & 28 & SAC & 29 \\
TD3 & 25 & V-trace & 8 \\
Non-RL developer memory & 40 & Total & 200 \\
\bottomrule
\end{tabular}
\end{table}

\FloatBarrier
\subsection{Controlled baselines and ablation evidence}

Table~\ref{tab:ablation} is the central ablation table. It distinguishes measured evidence from requested but uncollected variants. The uploaded evidence package supports a same-commit comparison between a deterministic control path and the full shadow/OPE configuration. It does not contain a pure static lexical/dense RAG replay, a pure deterministic-ranker-only replay, or intermediate controlled replays that isolate feedback normalization, delayed linking, and theory metadata one at a time. Those rows are therefore marked as not collected rather than filled with inferred numbers.

In the same-commit offline 200-case comparison, the deterministic control path and the full configuration both report 0.800 expected-decision accuracy, 0.000 hard-negative false-positive rate on the 80 hard-negative subset, and 0.200 failure/abstain rate. The measured difference is not accuracy; it is instrumentation and learnability. The deterministic control path has contextual-statistics update rate 0.000, whereas the full configuration has contextual-statistics update rate 0.605. This supports the narrower claim that the full architecture writes learning telemetry while preserving deterministic decision behavior on this benchmark, not that it outperforms the deterministic baseline in expected-decision accuracy.

\begin{table}[tbp]
\centering
\caption{Controlled baseline and ablation evidence matrix. FP rate is $1-$hard-negative suppression; FN is the observed failure/abstain rate.}
\label{tab:ablation}
\scriptsize
\setlength{\tabcolsep}{3pt}
\begin{tabular}{@{}lccccc rrrr l@{}}
\toprule
Variant & FB & DR & Bandit & OPE & Theory & Acc. & FP & FN & p95 ms & Status \\
\midrule
V0 static lexical/dense RAG & \xmark & \xmark & \xmark & \xmark & \xmark & -- & -- & -- & -- & not run \\
V1 deterministic ranker only & \xmark & \xmark & \xmark & \xmark & \xmark & -- & -- & -- & -- & not isolated \\
C0 deterministic control & \cmark & \cmark & logged & \xmark & \cmark & 0.800 & 0.000 & 0.200 & 66.5312 & measured \\
V2 + feedback norm & \cmark & \xmark & \xmark & \xmark & \xmark & -- & -- & -- & -- & not run \\
V3 + delayed link & \cmark & \cmark & \xmark & \xmark & \xmark & -- & -- & -- & -- & not run \\
V4 + theory metadata & \cmark & \cmark & \xmark & \xmark & \cmark & -- & -- & -- & -- & not run \\
V5 full shadow/OPE & \cmark & \cmark & \cmark & reported & \cmark & 0.800 & 0.000 & 0.200 & 66.3396 & measured \\
\bottomrule
\end{tabular}
\vspace{0.4em}
\begin{minipage}{0.96\linewidth}
\footnotesize
The uploaded evidence supports C0 and V5 under a shared same-commit outcome. V0--V4 are included as required experimental baselines but are not filled with inferred metrics; they require new controlled replays that disable static RAG, feedback normalization, delayed linking, theory metadata, and shadow/OPE components one at a time.
\end{minipage}
\end{table}

\begin{figure}[tbp]
\centering
\includegraphics[width=0.84\linewidth]{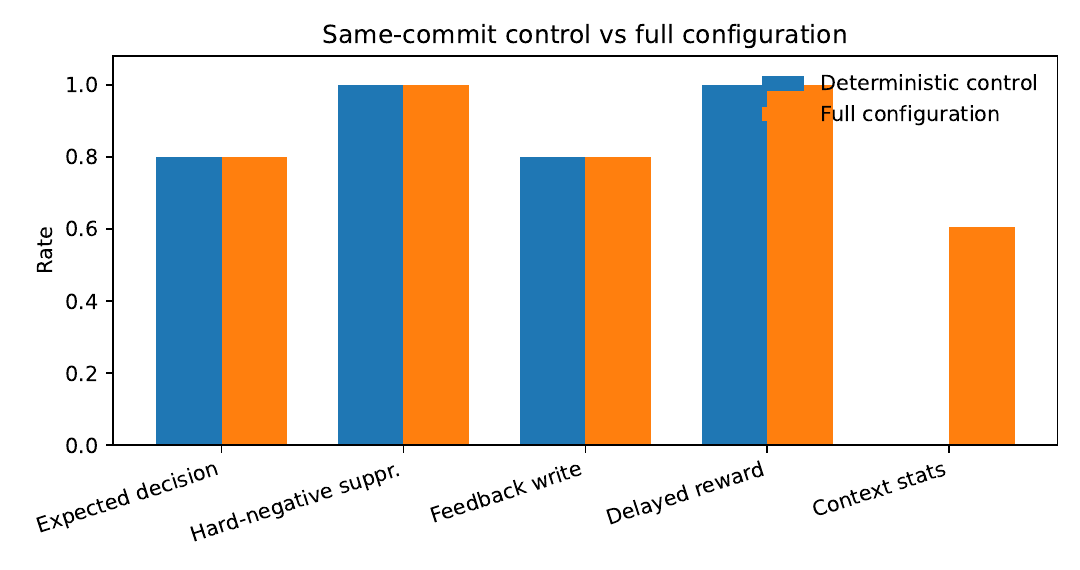}
\caption{Same-commit deterministic control versus full configuration. Decision metrics are unchanged, while the full configuration writes contextual-statistics telemetry.}
\label{fig:controlled-rates}
\end{figure}

\FloatBarrier
\subsection{Claim gate}

Table~\ref{tab:claim-gate} reports the claim gate as a first-class result. This table is intentionally stricter than the metric table: it states which claims are supported by the current evidence and which remain unsupported. The empirical contribution is strongest for executable architecture and component telemetry. Retrieval improvements and hard-negative improvements are only partially supported because patch-replay deltas are favorable, but the controlled same-commit comparison does not show a full-method accuracy advantage and the intermediate ablations are missing. Learned-policy deployment, official-client MCP interoperability, and end-to-end coding improvement are not supported by the current evidence.

\begin{table}[H]
\centering
\caption{Claim gate. Supported claims are limited to the uploaded evidence.}
\label{tab:claim-gate}
\footnotesize
\begin{tabularx}{\linewidth}{@{}p{0.20\linewidth} p{0.31\linewidth} >{\raggedright\arraybackslash}p{0.15\linewidth} Y@{}}
\toprule
Claim type & Required evidence & Status & Interpretation \\
\midrule
Architecture executable & Static, dynamic, and live tool evidence & Supported & Static 11/11, dynamic 10/10, and 200-case local-stdio live tool execution completed. \\
Retrieval decision improves & Accuracy, CI, and controlled ablation & Partially supported & Patch-replay deltas are favorable; same-commit control and full both have 0.800 expected-decision accuracy; V0--V4 are absent. \\
Hard-negative suppression improves & Paired hard-negative metric and CI & Partially supported & Patch-replay suppression changed from 0.975 to 1.000; same-commit control and full both report 1.000. \\
Learned policy deployable & OPE lower bound, support, and active canary & Not supported & Shadow/OPE telemetry exists, but active learned-policy canary is not demonstrated. \\
End-to-end coding improvement & Repository task patch success & Not supported & No SWE-bench or repository patch-success evaluation is included. \\
Official MCP interoperability & Official MCP client test & Not supported & Live evidence used local stdio JSON-RPC fallback because the official client module was unavailable. \\
\bottomrule
\end{tabularx}
\end{table}

\FloatBarrier
\subsection{Infrastructure and telemetry completion}

Infrastructure-level checks completed. Static validation passed 11/11 checks and dynamic integration passed 10/10 cases. In the final 200-case offline full-configuration run, retrieval events, bandit decisions, delayed reward telemetry, MCP tool success, algorithm-family matching, and review-gate correctness were recorded at 100\%. Feedback write rate was 80\%, reflecting the 40 residual failed rows in which the expected decision was \texttt{match} but the system returned \texttt{abstain}; no feedback was written for those rows. Contextual statistics update rate was 60.5\%, because learning updates occur only for feedback-bearing events rather than all retrievals.

\subsection{Patch-replay audit}

Table~\ref{tab:patch-replay} reports the patch-replay audit. This table answers a different question from Table~\ref{tab:ablation}: it asks whether the implementation revision changed measured behavior on paired cases. Several raw deltas are favorable, but the evidence package marks them as unconfirmed for broad improvement claims. The safest wording is metric-specific: offline pass rate changed from 71.5\% to 80.0\%; hard-negative suppression changed from 97.5\% to 100.0\%; online and live recall deltas moved from negative values to 0.0; contextual statistics update rate and live \texttt{issue\_match} success were unchanged. The claim gate reports no metric as fully improved and marks latency as regressed.

\begin{table}[tbp]
\centering
\caption{Patch-replay audit. Favorable raw deltas are retained for engineering transparency, not controlled method-superiority claims.}
\label{tab:patch-replay}
\scriptsize
\setlength{\tabcolsep}{4pt}
\begin{tabular}{@{}lrrrrl@{}}
\toprule
Metric & Earlier & Revised & Delta & $n$ & Gate status \\
\midrule
Offline pass & 0.715 & 0.800 & 0.085 & 200 & favorable, unconfirmed \\
Offline decision acc. & 0.790 & 0.800 & 0.010 & 200 & favorable, unconfirmed \\
Offline HN suppression & 0.975 & 1.000 & 0.025 & 200 & favorable, unconfirmed \\
HN subset suppression & 0.975 & 1.000 & 0.025 & 80 & favorable, unconfirmed \\
Online recall delta & -0.065 & 0.000 & 0.065 & 200 & favorable, unconfirmed \\
Live recall delta & -0.075 & 0.000 & 0.075 & 200 & favorable, unconfirmed \\
Online stats update & 0.605 & 0.605 & 0.000 & 200 & unchanged \\
Live match success & 1.000 & 1.000 & 0.000 & 200 & unchanged \\
Offline p95 latency (ms) & 60.5755 & 66.3396 & 5.7641 & 200 & regressed \\
Live p95 latency (ms) & 64.56045 & 67.4446 & 2.88415 & 200 & regressed \\
\bottomrule
\end{tabular}
\end{table}

\begin{figure}[tbp]
\centering
\includegraphics[width=0.84\linewidth]{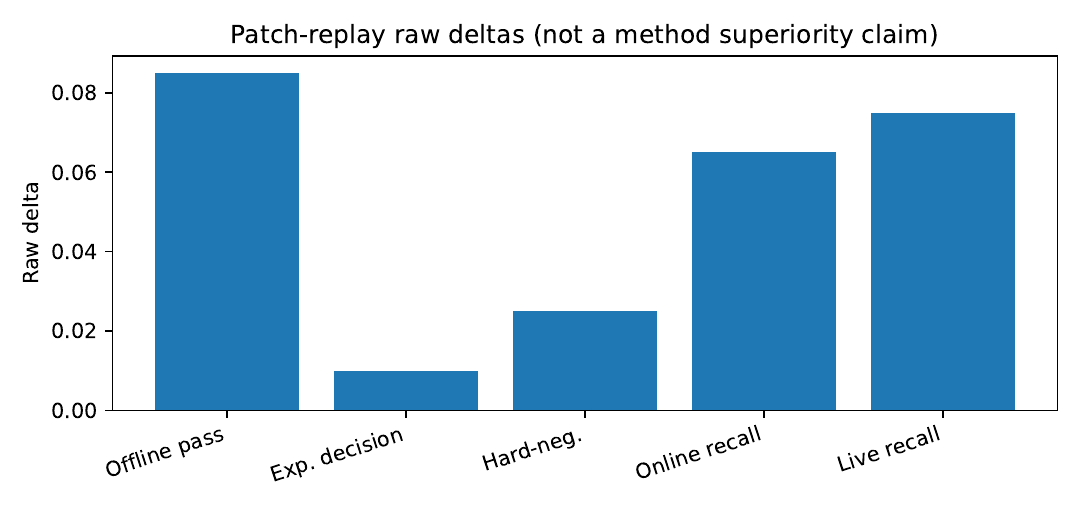}
\caption{Patch-replay raw deltas. These deltas are retained for engineering transparency but are not treated as controlled ablation evidence.}
\label{fig:patch-replay}
\end{figure}

\FloatBarrier
\subsection{Latency and operational cost}

Latency is the clearest operational cost. The same-commit offline issue-match p95 is similar between deterministic control and full configuration (66.5312 ms versus 66.3396 ms), but the live full configuration increases live p95 issue-match latency from 65.94575 ms to 67.4446 ms and live total-case p95 from 447.04895 ms to 852.45485 ms. Table~\ref{tab:latency-breakdown} gives the available tool-level breakdown. The current trace does not separately time query normalization, candidate retrieval, deterministic scoring, shadow scoring, and telemetry persistence inside \texttt{issue\_match}; those internal components must be instrumented separately before a causal latency decomposition can be claimed.

\begin{table}[tbp]
\centering
\caption{Live tool-level latency breakdown. Internal subcomponents of \texttt{issue\_match} are not separately instrumented in the current trace.}
\label{tab:latency-breakdown}
\scriptsize
\begin{tabularx}{\linewidth}{lrrrrY}
\toprule
Component & Full mean & Full p95 & Control mean & Control p95 & Notes \\
\midrule
Query normalization & -- & -- & -- & -- & Included in \texttt{issue\_match}; not separately timed. \\
Candidate retrieval & -- & -- & -- & -- & Included in \texttt{issue\_match}; not separately timed. \\
Deterministic scoring & -- & -- & -- & -- & Included in \texttt{issue\_match}; not separately timed. \\
Shadow bandit scoring & -- & -- & -- & -- & Included in \texttt{issue\_match}; not separately timed. \\
Telemetry persistence & -- & -- & -- & -- & Included in \texttt{issue\_match} and feedback writes; not separately timed. \\
\texttt{issue\_match} composite & 54.2339 & 67.4446 & 54.1056 & 65.9458 & Composite matching path. \\
\texttt{issue\_feedback} write & 0.7492 & 0.9083 & 0.6952 & 0.8606 & Explicit feedback tool. \\
Delayed link write & 6.4866 & 16.7161 & 5.7290 & 10.9020 & \texttt{issue\_record\_resolution} delayed path. \\
OPE/reporting proxy & 312.3108 & 692.8702 & 138.3234 & 310.8089 & \texttt{issue\_metrics} timing; proxy for reporting overhead, not pure OPE compute. \\
Health check & 20.0800 & 127.4320 & 19.5666 & 125.1120 & \texttt{issue\_health}. \\
Total live case & 445.0880 & 852.4549 & 270.1693 & 447.0490 & End-to-end live case timing. \\
\bottomrule
\end{tabularx}
\end{table}

\begin{figure}[tbp]
\centering
\includegraphics[width=0.84\linewidth]{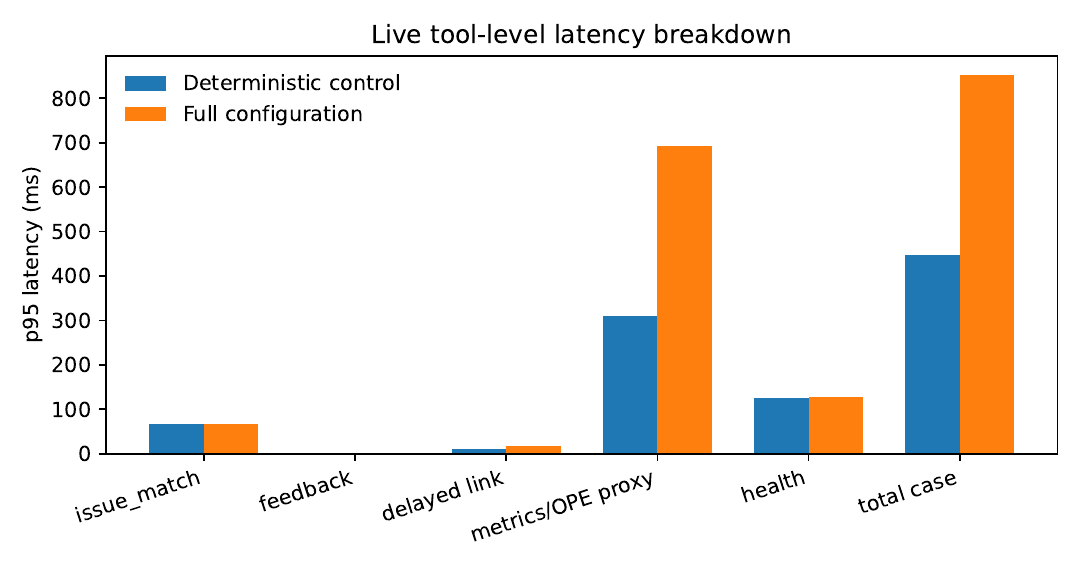}
\caption{Live tool-level p95 latency. The largest full-configuration increase is in metrics/reporting and total-case time, while issue-match p95 increases modestly.}
\label{fig:latency-breakdown}
\end{figure}

\FloatBarrier
\subsection{Residual failure family}

The residual failures in the revised implementation were not distributed across RL algorithm families. All 40 failures were in non-RL, low-risk developer-memory cases. They were evenly split across four bug families: SQLite path handling on Windows mounts, feedback aliases not canonicalized in the expected path, missing project scope, and command/path mismatch. In each family the expected decision was \texttt{match}, but the system returned \texttt{abstain}; feedback was therefore not written for those rows. Table~\ref{tab:residual} reports the family counts.

\begin{table}[tbp]
\centering
\caption{Residual failures in the revised implementation. All failures are non-RL developer-memory cases.}
\label{tab:residual}
\small
\begin{tabularx}{\linewidth}{lrlY}
\toprule
Residual bug family & Failed cases & Algorithm family & Observed behavior \\
\midrule
\texttt{sqlite\_path\_on\_windows\_mount} & 10 & non-RL & Expected \texttt{match}; returned \texttt{abstain}. \\
\texttt{feedback\_alias\_not\_canonical} & 10 & non-RL & Expected \texttt{match}; returned \texttt{abstain}. \\
\texttt{missing\_project\_scope} & 10 & non-RL & Expected \texttt{match}; returned \texttt{abstain}. \\
\texttt{command\_path\_mismatch} & 10 & non-RL & Expected \texttt{match}; returned \texttt{abstain}. \\
\midrule
Total & 40 & non-RL & One or more case assertions failed. \\
\bottomrule
\end{tabularx}
\end{table}

The residual distribution indicates that the current implementation is stronger on curated RL/control obligations than on general developer-memory normalization, especially path, scope, and alias compatibility. Therefore, the system should not yet be presented as a mature general-purpose developer-memory layer.

\FloatBarrier
\section{Discussion}

\sys\ turns developer memory into a governed decision layer for RL coding agents. Its retrieval path keeps a deterministic ranker as the deployed baseline, while the learning path records shadow contextual-bandit scores, propensities, normalized rewards, delayed links, OPE summaries, and review-gated theory-to-code metadata. This design deliberately sits between passive retrieval and unconstrained online RL. It learns from feedback, but it does not allow unverified learned ranking to control live memory selection; it uses developer memory, but it treats every match as an auditable decision with a safety boundary.

The results support this architectural claim at the component level, not as a broad method-superiority claim. Static validation passed all 11 checks, dynamic integration passed all 10 cases, live tool calls completed under local stdio JSON-RPC fallback, and the final 200-case runs produced retrieval events, bandit telemetry, delayed-reward records, algorithm-family matching, and review-gate telemetry. The same-commit deterministic control and the full shadow/OPE configuration both report 0.800 expected-decision accuracy, 1.000 hard-negative suppression, and 0.200 failure/abstain rate. The measured distinction is that the full configuration writes contextual-statistics and OPE/shadow telemetry, whereas the deterministic control does not demonstrate an accuracy disadvantage or advantage on this benchmark. The patch-replay audit is still useful engineering evidence because it records favorable raw deltas in pass rate, hard-negative suppression, and recall-delta metrics; however, those deltas are explicitly separated from the controlled baseline table. The latency regressions and the 40 residual non-RL failures are not secondary details; they define the present engineering boundary of the system.

Compared with static RAG and vector memory, the main advantage of conventional retrieval is simplicity. A standard RAG pipeline can retrieve semantically similar documents and prepend them to a prompt with relatively low architectural overhead \citep{lewis2020rag}. Repository-aware graph retrieval further improves how code context is organized across files and dependencies \citep{fedorov2025gracg}. Utility-driven episodic-memory systems such as MemRL provide a closer learning-based contrast because they update memory from runtime utility rather than only semantic similarity \citep{zhang2026memrl}. These methods are attractive for documentation lookup, code search, and single-turn context expansion, and they motivate the broader shift from passive retrieval to memory control. Their limitation in this setting is that semantic or graph proximity is not the same as repair utility. A memory can be lexically close to an RL bug while violating the relevant Bellman target, bootstrap mask, loss term, or validation tier. \sys\ adds a feedback-bearing decision contract around retrieval so that accepted, rejected, delayed, false-positive, and review-gated events become part of the memory state. The cost is visible in the latency regression and in the need to maintain richer metadata.

Compared with long-context agents and general long-term memory systems, \sys\ makes a different trade-off. Long-context agents are useful when all relevant evidence fits in the effective context and the task is short enough that repeated full-context prompting remains economical. Long-term assistant memory systems such as MemoryBank and evaluation suites such as LongMemEval show why persistent memory is valuable, but they also emphasize temporal updates, abstention, and multi-session reasoning as separate capabilities rather than solved consequences of a larger window \citep{zhong2024memorybank,wu2024longmemeval}. The lost-in-the-middle effect reinforces that more tokens do not guarantee correct use of buried evidence \citep{liu2024lostmiddle}. \sys\ complements, rather than replaces, context windows by storing compact validated records with feedback state and audit metadata. The residual abstentions in path, scope, and alias cases show the downside of this choice: external memory can fail conservatively when its specificity predicates are too strict.

Compared with self-correction and memory-augmented debugging systems, the contribution is located in the memory infrastructure rather than in the generator policy. Reflexion and Self-Refine demonstrate that agents can benefit from iterative feedback and self-generated revision traces \citep{shinn2023reflexion,madaan2023selfrefine}. Interactive code-debugging work with human feedback and memory shows the value of combining repair attempts with stored developer feedback \citep{krishnamoorthy2025debugging}. SCoRe-style training uses online RL to improve self-correction behavior in the model policy \citep{kumar2024selfcorrect}. These approaches are stronger candidates for improving repair behavior directly, but they do not by themselves guarantee that a persistent memory store will avoid stale retrieval, false-positive reinforcement, or unsafe RL/control promotion. \sys\ instead treats memory selection as the object of control: the retrieved record, feedback event, delayed resolution, and governance state are all explicit artifacts.

The contextual-bandit component is intentionally modest. Learning to rank can be formulated as a contextual bandit because the system chooses among candidate memories and observes feedback on the chosen item \citep{li2010contextual,gampa2021banditrank}. Memory-augmented contextual bandits are also appropriate when interactions are sparse and history-dependent \citep{shen2018memorybandit}. A full MDP formulation would be more expressive, but it would also introduce harder credit assignment and unsafe exploration. Shadow execution and continuous safety assessment provide a safer transition path because candidate learned behavior can be evaluated beside a baseline before activation \citep{gassert2024shadow,guissouma2023shadow}. OPE methods, including doubly robust and counterfactual logged-bandit estimators, provide the statistical machinery for evaluating target policies from historical data \citep{dudik2011doubly,swaminathan2015counterfactual}. Conservative offline RL further motivates lower-bound and support-aware promotion criteria \citep{kumar2020cql,yu2021combo}. The limitation is equally clear: a contextual bandit estimates memory-selection utility, not the full downstream trajectory of code editing, testing, and developer acceptance.

The RL-specific metadata is the main reason the system should not be described as generic RAG with extra logging. RL code correctness depends on algorithmic obligations, not only on strings. Reproducibility, hyperparameter, and sensitivity studies show that small variations in RL code and evaluation practice can materially alter outcomes \citep{henderson2018deep,eimer2023hyperparameters,adkins2024sensitivity}. Empirical RL-bug studies show that agent-environment interaction and algorithmic logic are central failure sources \citep{song2026bugs}. Theory-to-code memory records therefore store algorithm family, update-equation obligations, runtime stage, validation tier, audit findings, and code anchors. This helps prevent a retrieved fix from being promoted solely because it looks similar. It does not prove mathematical correctness, and the paper should not imply such a proof. It provides traceability and reviewability, which are narrower and more defensible claims.

The safety posture is conservative because persistent developer memory can become a liability. Code memorization work demonstrates that code models can expose sensitive artifacts \citep{alkaswan2024traces}; studies of code helpers identify credential and security risks \citep{sheggam2024security}; long-term assistant memory raises retention and access-control concerns \citep{lee2024ethicalmemory}; and agentic systems face prompt-injection and data-manipulation threats \citep{khan2024agenticsecurity}. Local-first storage, redaction boundaries, canonical feedback, idempotent delayed links, false-positive suppression, review-gated RL/control promotion, and OPE-gated rollout reduce the authority of unverified memory. They do not guarantee absolute privacy, immunity to prompt injection, or complete protection from memory poisoning. The architecture is therefore best viewed as risk reduction through auditability and constrained authority.

The evaluation design is intentionally conservative. It isolates the memory layer rather than measuring arbitrary developer productivity or broad SWE-bench performance. This avoids conflating the base coding agent's generation ability with the memory-control system's contribution. Benchmarks such as LoCoBench-Agent make clear that realistic software-engineering evaluation should include multi-turn tool use and long-context repository interaction \citep{qiu2025locobenchagent}; the present study does not yet claim that level of end-to-end evidence. The trade-off is that the strongest results are component-level: tool-chain execution, telemetry completion, hard-negative suppression, recall-delta behavior, latency regression, and residual family characterization. The missing V0--V4 controlled ablations are therefore not cosmetic; they are the next experiments required to isolate static RAG, deterministic ranking, feedback normalization, delayed linking, and theory metadata under one shared outcome. Future work should expand beyond deterministic cases to real developer traces, official-client MCP interoperability, larger OPE-supported low-risk canaries, stronger path/scope/alias normalization, and end-to-end repository tasks where memory-control evidence can be connected to patch quality without overstating causality.

\section{Conclusion}

This paper presented \sys, a feedback-normalized, delayed-reward-aware, contextual-bandit-shadowed, OPE-gated, theory-to-code-traceable developer-memory architecture for RL coding agents. The system is implemented as a local-first MCP server with explicit tools for matching, feedback, and resolution recording. It preserves a deterministic ranker as the deployed baseline, learns only from bounded canonical reward events, links delayed verified resolutions to earlier retrieval events, and restricts active learned ranking through conservative OPE and safety predicates. RL/control memories are additionally constrained by validation-tier promotion and theory-to-code traceability.

The evaluation supports component-level conclusions. Static and dynamic infrastructure checks passed; offline, online-shadow, and live evaluations completed; the same-commit deterministic control and full configuration preserved the same expected-decision accuracy and hard-negative suppression; and the patch-replay audit recorded favorable raw deltas that are kept separate from controlled baseline claims. The evaluation also identifies constraints: the requested V0--V4 controlled ablations are not yet measured, no broad performance-gain claim is supported, active canary behavior is not established, p95 latency regressed, official-client MCP interoperability was not shown, and 40 residual non-RL developer-memory failures remained. Future work should collect larger real developer traces, execute official-client MCP evaluations, run controlled static-RAG and intermediate ablations, perform OPE-supported active canaries only on low-risk families, improve path/scope/alias coverage, and evaluate whether the memory-control layer improves end-to-end coding-agent outcomes on realistic repository tasks.

\section*{Acknowledgements}
The author acknowledges PythaLab and Yildiz Technical University for supporting this work. The implementation and supplementary artifacts are available at \url{https://github.com/PhiniteLab/rl-developer-memory}.

\bibliographystyle{plainnat}
\bibliography{references}

@inproceedings{lewis2020rag,
  title={Retrieval-Augmented Generation for Knowledge-Intensive NLP Tasks},
  author={Lewis, Patrick and Perez, Ethan and Piktus, Aleksandra and Petroni, Fabio and Karpukhin, Vladimir and Goyal, Naman and K{\"u}ttler, Heinrich and Lewis, Mike and Yih, Wen-tau and Rockt{\"a}schel, Tim and Riedel, Sebastian and Kiela, Douwe},
  booktitle={Advances in Neural Information Processing Systems},
  volume={33},
  pages={9459--9474},
  year={2020},
  doi={10.48550/arXiv.2005.11401},
  url={https://arxiv.org/abs/2005.11401}
}

@inproceedings{zhong2024memorybank,
  title={MemoryBank: Enhancing Large Language Models with Long-Term Memory},
  author={Zhong, Wanjun and Guo, Lianghong and Gao, Qiqi and Ye, He and Wang, Yanlin},
  booktitle={Proceedings of the AAAI Conference on Artificial Intelligence},
  volume={38},
  pages={19724--19731},
  year={2024},
  doi={10.1609/aaai.v38i17.29946},
  url={https://doi.org/10.1609/aaai.v38i17.29946}
}

@article{wu2024longmemeval,
  title={LongMemEval: Benchmarking Chat Assistants on Long-Term Interactive Memory},
  author={Wu, Di and Wang, Hongwei and Yu, Wenhao and Zhang, Yuwei and Chang, Kai-Wei and Yu, Dong},
  journal={arXiv preprint arXiv:2410.10813},
  year={2024},
  doi={10.48550/arXiv.2410.10813},
  url={https://arxiv.org/abs/2410.10813}
}

@article{liu2024lostmiddle,
  title={Lost in the Middle: How Language Models Use Long Contexts},
  author={Liu, Nelson F. and Lin, Kevin and Hewitt, John and Paranjape, Ashwin and Bevilacqua, Michele and Petroni, Fabio and Liang, Percy},
  journal={Transactions of the Association for Computational Linguistics},
  volume={12},
  pages={157--173},
  year={2024},
  doi={10.1162/tacl_a_00638},
  url={https://doi.org/10.1162/tacl_a_00638}
}

@inproceedings{jimenez2024swebench,
  title={{SWE}-bench: Can Language Models Resolve Real-World GitHub Issues?},
  author={Jimenez, Carlos E. and Yang, John and Wettig, Alexander and Yao, Shunyu and Pei, Kexin and Press, Ofir and Narasimhan, Karthik},
  booktitle={International Conference on Learning Representations},
  year={2024},
  doi={10.48550/arXiv.2310.06770},
  url={https://arxiv.org/abs/2310.06770},
  note={OpenReview: \url{https://openreview.net/forum?id=VTF8yNQM66}}
}

@inproceedings{yang2024sweagent,
  title={{SWE}-agent: Agent-Computer Interfaces Enable Automated Software Engineering},
  author={Yang, John and Jimenez, Carlos E. and Wettig, Alexander and Lieret, Kilian and Yao, Shunyu and Narasimhan, Karthik and Press, Ofir},
  booktitle={Advances in Neural Information Processing Systems},
  year={2024},
  doi={10.48550/arXiv.2405.15793},
  url={https://arxiv.org/abs/2405.15793}
}

@inproceedings{yao2023react,
  title={{ReAct}: Synergizing Reasoning and Acting in Language Models},
  author={Yao, Shunyu and Zhao, Jeffrey and Yu, Dian and Du, Nan and Shafran, Izhak and Narasimhan, Karthik and Cao, Yuan},
  booktitle={International Conference on Learning Representations},
  year={2023},
  doi={10.48550/arXiv.2210.03629},
  url={https://arxiv.org/abs/2210.03629},
  note={OpenReview: \url{https://openreview.net/forum?id=WE_vluYUL-X}}
}

@inproceedings{madaan2023selfrefine,
  title={Self-Refine: Iterative Refinement with Self-Feedback},
  author={Madaan, Aman and Tandon, Niket and Gupta, Prakhar and Hallinan, Skyler and Gao, Luyu and Wiegreffe, Sarah and Alon, Uri and Dziri, Nouha and Prabhumoye, Shrimai and Yang, Yiming and Welleck, Sean and Majumder, Bodhisattwa Prasad and Gupta, Shashank and Yazdanbakhsh, Amir and Clark, Peter},
  booktitle={Advances in Neural Information Processing Systems},
  volume={36},
  year={2023},
  doi={10.48550/arXiv.2303.17651},
  url={https://arxiv.org/abs/2303.17651}
}

@inproceedings{shinn2023reflexion,
  title={Reflexion: Language Agents with Verbal Reinforcement Learning},
  author={Shinn, Noah and Cassano, Federico and Berman, Edward and Gopinath, Ashwin and Narasimhan, Karthik and Yao, Shunyu},
  booktitle={Advances in Neural Information Processing Systems},
  year={2023},
  doi={10.48550/arXiv.2303.11366},
  url={https://arxiv.org/abs/2303.11366},
  note={OpenReview: \url{https://openreview.net/forum?id=vAElhFcKW6}}
}

@inproceedings{henderson2018deep,
  title={Deep Reinforcement Learning That Matters},
  author={Henderson, Peter and Islam, Riashat and Bachman, Philip and Pineau, Joelle and Precup, Doina and Meger, David},
  booktitle={Proceedings of the AAAI Conference on Artificial Intelligence},
  pages={3207--3214},
  year={2018},
  doi={10.48550/arXiv.1709.06560},
  url={https://arxiv.org/abs/1709.06560}
}

@inproceedings{eimer2023hyperparameters,
  title={Hyperparameters in Reinforcement Learning and How to Tune Them},
  author={Eimer, Theresa and Lindauer, Marius and Raileanu, Roberta},
  booktitle={Proceedings of the 40th International Conference on Machine Learning},
  series={Proceedings of Machine Learning Research},
  volume={202},
  pages={9104--9149},
  year={2023},
  doi={10.48550/arXiv.2306.01324},
  url={https://proceedings.mlr.press/v202/eimer23a.html}
}

@inproceedings{adkins2024sensitivity,
  title={A Method for Evaluating Hyperparameter Sensitivity in Reinforcement Learning},
  author={Adkins, Jacob and Bowling, Michael and White, Adam},
  booktitle={Advances in Neural Information Processing Systems},
  volume={37},
  pages={124820--124842},
  year={2024},
  doi={10.48550/arXiv.2412.07165},
  url={https://arxiv.org/abs/2412.07165}
}

@article{gu2025domaincode,
  title={On the Effectiveness of Large Language Models in Domain-Specific Code Generation},
  author={Gu, Xiaodong and Chen, Meng and Lin, Yalan and Hu, Yuhan and Zhang, Hongyu and Wan, Chengcheng and Wei, Zhao and Xu, Yong and Wang, Juhong},
  journal={ACM Transactions on Software Engineering and Methodology},
  volume={34},
  number={3},
  pages={78:1--78:28},
  year={2025},
  doi={10.1145/3697012},
  url={https://doi.org/10.1145/3697012}
}

@inproceedings{fedorov2025gracg,
  title={{GRACG}: Graph Retrieval Augmented Code Generation},
  author={Fedorov, Konstantin and Zarubin, Boris and Ivanov, Vladimir},
  booktitle={2025 40th IEEE/ACM International Conference on Automated Software Engineering Workshops},
  pages={291--298},
  year={2025},
  doi={10.1109/ASEW67777.2025.00060},
  url={https://doi.org/10.1109/ASEW67777.2025.00060}
}

@article{song2026bugs,
  title={Investigating the Bugs in Reinforcement Learning Programs: Insights from Stack Overflow and GitHub},
  author={Song, Jiayin and Li, Yike and Tian, Yunzhe and Ma, Haoxuan and Li, Honglei and Zuo, Jie and Liu, Jiqiang and Niu, Wenjia},
  journal={Automated Software Engineering},
  volume={33},
  number={1},
  pages={9},
  year={2026},
  doi={10.1007/s10515-025-00555-z},
  url={https://doi.org/10.1007/s10515-025-00555-z}
}

@inproceedings{krishnamoorthy2025debugging,
  title={Multi-Agent Reinforcement Learning for Interactive Code Debugging with Human Feedback and Memory},
  author={Krishnamoorthy, Anjana and Ivatury, Kartik and Ahmadnia, Benyamin},
  booktitle={Proceedings of the 15th International Conference on Recent Advances in Natural Language Processing - Natural Language Processing in the Generative AI Era},
  pages={595--603},
  address={Varna, Bulgaria},
  publisher={INCOMA Ltd., Shoumen, Bulgaria},
  year={2025},
  url={https://aclanthology.org/2025.ranlp-1.70/}
}

@article{kumar2024selfcorrect,
  title={Training Language Models to Self-Correct via Reinforcement Learning},
  author={Kumar, Aviral and Zhuang, Vincent and Agarwal, Rishabh and Su, Yi and Co-Reyes, John D. and Singh, Avi and Baumli, Kate and Iqbal, Shariq and Bishop, Colton and Roelofs, Rebecca and Zhang, Lei M. and McKinney, Kay and Shrivastava, Disha and Paduraru, Cosmin and Tucker, George and Precup, Doina and Behbahani, Feryal and Faust, Aleksandra},
  journal={arXiv preprint arXiv:2409.12917},
  year={2024},
  doi={10.48550/arXiv.2409.12917},
  url={https://arxiv.org/abs/2409.12917}
}

@inproceedings{alkaswan2024traces,
  title={Traces of Memorisation in Large Language Models for Code},
  author={Al-Kaswan, Ali and Izadi, Maliheh and Van Deursen, Arie},
  booktitle={Proceedings of the IEEE/ACM 46th International Conference on Software Engineering},
  pages={1--12},
  year={2024},
  doi={10.1145/3597503.3639133},
  url={https://doi.org/10.1145/3597503.3639133}
}

@inproceedings{sheggam2024security,
  title={Exploring Security Risks and Mitigation Strategies in AI Code Helpers},
  author={Sheggam, Harshith and Zhang, Xiaowen},
  booktitle={2024 IEEE Long Island Systems, Applications and Technology Conference},
  pages={1--6},
  year={2024},
  doi={10.1109/LISAT63094.2024.10807934},
  url={https://doi.org/10.1109/LISAT63094.2024.10807934}
}

@article{lee2024ethicalmemory,
  title={Towards Ethical Personal AI Applications: Practical Considerations for AI Assistants with Long-Term Memory},
  author={Lee, Eunji},
  journal={arXiv preprint arXiv:2409.11192},
  year={2024},
  doi={10.48550/arXiv.2409.11192},
  url={https://arxiv.org/abs/2409.11192}
}

@article{khan2024agenticsecurity,
  title={Security Threats in Agentic AI System},
  author={Khan, Raihan and Sarkar, Soumik and Mahata, Shubha Kanti and Jose, Erwin},
  journal={arXiv preprint arXiv:2410.14728},
  year={2024},
  doi={10.48550/arXiv.2410.14728},
  url={https://arxiv.org/abs/2410.14728}
}

@inproceedings{li2010contextual,
  title={A Contextual-Bandit Approach to Personalized News Article Recommendation},
  author={Li, Lihong and Chu, Wei and Langford, John and Schapire, Robert E.},
  booktitle={Proceedings of the 19th International Conference on World Wide Web},
  pages={661--670},
  year={2010},
  doi={10.1145/1772690.1772758},
  url={https://doi.org/10.1145/1772690.1772758}
}

@inproceedings{gampa2021banditrank,
  title={BanditRank: Learning to Rank Using Contextual Bandits},
  author={Gampa, Phanideep and Fujita, Sumio},
  booktitle={Pacific-Asia Conference on Knowledge Discovery and Data Mining},
  pages={259--271},
  publisher={Springer},
  year={2021},
  doi={10.1007/978-3-030-75768-7_21},
  url={https://doi.org/10.1007/978-3-030-75768-7_21}
}

@inproceedings{shen2018memorybandit,
  title={Interactive Recommendation via Deep Neural Memory Augmented Contextual Bandits},
  author={Shen, Yilin and Deng, Yang and Ray, Avik and Jin, Hongxia},
  booktitle={Proceedings of the 12th ACM Conference on Recommender Systems},
  pages={122--130},
  year={2018},
  doi={10.1145/3240323.3240344},
  url={https://doi.org/10.1145/3240323.3240344}
}

@inproceedings{dudik2011doubly,
  title={Doubly Robust Policy Evaluation and Learning},
  author={Dud{\i}k, Miroslav and Langford, John and Li, Lihong},
  booktitle={Proceedings of the 28th International Conference on Machine Learning},
  pages={1097--1104},
  year={2011},
  doi={10.48550/arXiv.1103.4601},
  url={https://arxiv.org/abs/1103.4601}
}

@inproceedings{swaminathan2015counterfactual,
  title={Counterfactual Risk Minimization: Learning from Logged Bandit Feedback},
  author={Swaminathan, Adith and Joachims, Thorsten},
  booktitle={Proceedings of the 32nd International Conference on Machine Learning},
  series={Proceedings of Machine Learning Research},
  volume={37},
  pages={814--823},
  year={2015},
  url={https://proceedings.mlr.press/v37/swaminathan15.html}
}

@inproceedings{kumar2020cql,
  title={Conservative Q-Learning for Offline Reinforcement Learning},
  author={Kumar, Aviral and Zhou, Aurick and Tucker, George and Levine, Sergey},
  booktitle={Advances in Neural Information Processing Systems},
  volume={33},
  pages={1179--1191},
  year={2020},
  doi={10.48550/arXiv.2006.04779},
  url={https://arxiv.org/abs/2006.04779}
}

@inproceedings{yu2021combo,
  title={{COMBO}: Conservative Offline Model-Based Policy Optimization},
  author={Yu, Tianhe and Kumar, Aviral and Rafailov, Rafael and Rajeswaran, Aravind and Levine, Sergey and Finn, Chelsea},
  booktitle={Advances in Neural Information Processing Systems},
  volume={34},
  pages={28954--28967},
  year={2021},
  doi={10.48550/arXiv.2102.08363},
  url={https://arxiv.org/abs/2102.08363}
}

@article{gassert2024shadow,
  title={Stepping Out of the Shadows: Reinforcement Learning in Shadow Mode},
  author={Gassert, Patrick and Althoff, Matthias},
  journal={arXiv preprint arXiv:2410.23419},
  year={2024},
  doi={10.48550/arXiv.2410.23419},
  url={https://arxiv.org/abs/2410.23419}
}

@inproceedings{guissouma2023shadow,
  title={Continuous Safety Assessment of Updated Supervised Learning Models in Shadow Mode},
  author={Guissouma, Hatem and Zink, Markus and Sax, Eric},
  booktitle={2023 IEEE 20th International Conference on Software Architecture Companion},
  pages={301--308},
  year={2023},
  doi={10.1109/ICSA-C57050.2023.00069},
  url={https://doi.org/10.1109/ICSA-C57050.2023.00069}
}

@misc{mcpdocs2026,
  title={Model Context Protocol Specification},
  author={{Model Context Protocol}},
  year={2026},
  url={https://modelcontextprotocol.io/specification/2025-11-25},
  note={Accessed May 2, 2026}
}

@inproceedings{lyle2024normalization,
  title={Normalization and Effective Learning Rates in Reinforcement Learning},
  author={Lyle, Clare and Zheng, Zeyu and Khetarpal, Khimya and Martens, James and van Hasselt, Hado and Pascanu, Razvan and Dabney, Will},
  booktitle={Advances in Neural Information Processing Systems},
  volume={37},
  pages={106440--106473},
  year={2024},
  doi={10.48550/arXiv.2407.01800},
  url={https://arxiv.org/abs/2407.01800}
}

@inproceedings{qu2025latentreward,
  title={Latent Reward: LLM-Empowered Credit Assignment in Episodic Reinforcement Learning},
  author={Qu, Yun and Jiang, Yuhang and Wang, Boyuan and Mao, Yixiu and Wang, Cheems and Liu, Chang and Ji, Xiangyang},
  booktitle={Proceedings of the AAAI Conference on Artificial Intelligence},
  volume={39},
  pages={20095--20103},
  year={2025},
  doi={10.1609/aaai.v39i19.34213},
  url={https://doi.org/10.1609/aaai.v39i19.34213}
}

@article{qiu2025locobenchagent,
  title={{LoCoBench-Agent}: An Interactive Benchmark for LLM Agents in Long-Context Software Engineering},
  author={Qiu, Jielin and Liu, Zuxin and Liu, Zhiwei and Murthy, Rithesh and Zhang, Jianguo and Chen, Haolin and Wang, Shiyu and Zhu, Ming and Yang, Liangwei and Tan, Juntao and Ram, Roshan and Prabhakar, Akshara and Awalgaonkar, Tulika and Chen, Zixiang and Cen, Zhepeng and Qian, Cheng and Heinecke, Shelby and Yao, Weiran and Savarese, Silvio and Xiong, Caiming and Wang, Huan},
  journal={arXiv preprint arXiv:2511.13998},
  year={2025},
  doi={10.48550/arXiv.2511.13998},
  url={https://arxiv.org/abs/2511.13998}
}

@article{zhang2026memrl,
  title={{MemRL}: Self-Evolving Agents via Runtime Reinforcement Learning on Episodic Memory},
  author={Zhang, Shengtao and Wang, Jiaqian and Zhou, Ruiwen and Liao, Junwei and Feng, Yuchen and Li, Zhiyu and Xiong, Feiyu and Qi, Yutao and Tang, Bo and Zhang, Weinan and Wen, Muning},
  journal={arXiv preprint arXiv:2601.03192},
  year={2026},
  doi={10.48550/arXiv.2601.03192},
  url={https://arxiv.org/abs/2601.03192}
}

\end{document}